\documentclass[final,5p,times,twocolumn]{elsarticle}

\usepackage{amssymb}
\usepackage{multirow}
\usepackage{xcolor}
\usepackage{soul}
\usepackage{upgreek}
\newcommand{\n}{{\rm n}}
\newcommand{\Hg}{{\rm Hg}}

\begin{document}

\begin{frontmatter}

\title{A measurement of the neutron to $^{199}$Hg magnetic moment ratio}

\author[ETH,PSI,JUH]{S.~Afach}
\author[RAL]{C.~A.~Baker}
\author[LPC]{G.~Ban}
\author[PSI]{G.~Bison}
\author[JUC]{K.~Bodek} 
\author[PTB]{M.~Burghoff} 
\author[PSI]{Z.~Chowdhuri}
\author[PSI]{M.~Daum}       
\author[ETH,PSI]{M.~Fertl\fnref{MF}}		
\author[ETH,PSI]{B.~Franke\fnref{BF}}
\author[ILL]{P.~Geltenbort}  	
\author[RAL,SUSSEX]{K.~Green}
\author[RAL,SUSSEX]{M.~G.~D.~van~der~Grinten} 
\author[UNIFR]{Z.~Grujic} 
\author[SUSSEX]{P.~G.~Harris}
\author[PGUM]{W. Heil}         	
\author[PSI,LPC]{V.~H\'elaine}   	
\author[PSI]{R.~Henneck}
\author[ETH,PSI]{M.~Horras}       	
\author[RAL]{P.~Iaydjiev \fnref{INRNE}}
\author[RAL]{S.~N.~Ivanov \fnref{PNPI}}
\author[UNIFR]{M.~Kasprzak}
\author[LPSC]{Y.~Kerma\"{i}dic}
\author[ETH,PSI]{K.~Kirch}
\author[PSI]{A.~Knecht}
\author[UNIFR,PGUM]{H.-C.~Koch}
\author[ETH]{J.~Krempel}      	
\author[PSI,JUC]{M. Ku\'zniak \fnref{MK}}
\author[PSI]{B.~Lauss}        	
\author[LPC]{T.~Lefort}       	
\author[LPC]{Y.~Lemi\`ere}
\author[PSI]{A.~Mtchedlishvili}     
\author[LPC]{O.~Naviliat-Cuncic\fnref{ONC}}
\author[SUSSEX]{J.~M.~Pendlebury}
\author[JUC]{M.~Perkowski}
\author[PSI,LPC]{E.~Pierre}
\author[ETH]{F.~M.~Piegsa}	
\author[LPSC]{G.~Pignol}         \ead{pignol@lpsc.in2p3.fr}
\author[KULEUVEN]{P.~N.~Prashanth}
\author[LPC]{G.~Qu\'em\'ener} 	
\author[LPSC]{D.~Rebreyend}
\author[PSI]{D.~Ries}    	
\author[CSNSM]{S.~Roccia}
\author[PSI]{P.~Schmidt-Wellenburg}
\author[PTB]{A.~Schnabel}  
\author[KULEUVEN]{N.~Severijns}    
\author[SUSSEX]{D.~Shiers}       	
\author[SUSSEX]{K.~F.~Smith\fnref{KFS}}
\author[PTB]{J.~Voigt}		
\author[UNIFR]{A. Weis} 
\author[ETH,JUC]{G.~Wyszynski}        	
\author[JUC]{J.~Zejma}        
\author[ETH,PSI,IKC]{J.~Zenner}		
\author[PSI]{G.~Zsigmond}     	

\address[ETH]{ETH Z\"urich, Institute for Particle Physics, CH-8093 Z\"urich, Switzerland}
\address[PSI]{Paul Scherrer Institute (PSI), CH--5232 Villigen-PSI, Switzerland}
\address[JUH]{Hans Berger Department of Neurology, Jena University Hospital, D-07747 Jena, Germany}
\address[RAL]{Rutherford Appleton Laboratory, Chilton, Didcot, Oxon OX11 0QX, United Kingdom}
\address[LPC]{LPC Caen, ENSICAEN, Universit\'e de Caen, CNRS/IN2P3, Caen, France}
\address[JUC]{Marian Smoluchowski Institute of Physics, Jagiellonian University, 30--059 Cracow, Poland}
\address[PTB]{Physikalisch Technische Bundesanstalt, Berlin, Germany}
\address[ILL]{Institut Laue--Langevin, Grenoble, France}
\address[SUSSEX]{Department of Physics and Astronomy, University of Sussex, Falmer, Brighton BN1 9QH, United Kingdom}
\address[UNIFR]{Physics Department, University of Fribourg, CH-1700 Fribourg, Switzerland}
\address[PGUM]{Institut f\"ur Physik, Johannes--Gutenberg--Universit\"at, D--55128 Mainz, Germany}
\address[LPSC]{LPSC, Universit\'e  Grenoble Alpes, CNRS/IN2P3, Grenoble, France}
\address[KULEUVEN]{Instituut voor Kern-- en Stralingsfysica, Katholieke~Universiteit~Leuven, B--3001 Leuven, Belgium}
\address[CSNSM]{CSNSM, Universit\'e Paris Sud, CNRS/IN2P3, Orsay Campus, France}
\address[IKC]{Institut f\"{u}r Kernchemie, Johannes-Gutenberg-Universitat, D-55128 Mainz, Germany}

\fntext[MF]{Now at University of Washington, Seattle WA, USA.}
\fntext[BF]{Now at Max-Planck-Institute of Quantum Optics, Garching, Germany.}
\fntext[INRNE]{On leave from INRNE, Sofia, Bulgaria.}
\fntext[PNPI]{On leave from PNPI, St. Petersburg, Russia.}
\fntext[MK]{Now at Queen's University, Kingston ON, Canada.}
\fntext[ONC]{Now at Michigan State University, East-Lansing, USA.}
\fntext[KFS]{Deceased.}

\begin{abstract}
The neutron gyromagnetic ratio has been measured relative to that of the $^{199}$Hg atom 
with an uncertainty of $0.8$~ppm. 
We employed an apparatus where ultracold neutrons and mercury atoms are stored in the same volume and 
report the result $\gamma_\n/\gamma_\Hg = 3.8424574(30)$. 
\end{abstract}


\end{frontmatter}


\section{Introduction}
\label{Introduction}

After Chadwick's discovery of the neutron in 1932, it became clear that nuclei are made out of protons and neutrons. 
In this picture the neutron had to bear a nonzero magnetic moment in order to account for the magnetic moments of nuclei. 
The observation that the neutron, an electrically neutral particle, has a nonzero magnetic moment 
is in conflict with the prediction of the relativistic Dirac equation valid for elementary spin 1/2 particles. 
This fact was an early indication of the existence of a sub-structure for neutrons and protons. 

The first direct measurement of the neutron magnetic moment was reported by Alvarez and Bloch in 1940 \cite{Alvarez1940} with an uncertainty of one percent. 
The frequency $f_\n$ of the neutron spin precession in a magnetic field $B_{\rm 0}$ was measured using the Rabi resonance method, 
from which the gyromagnetic ratio $\gamma_\n = 2 \pi f_\n / B_{\rm 0}$ and the magnetic moment $\mu_\n = \hbar/2 \, \gamma_\n$ were extracted. 
The precision of this method was then improved by using the proton magnetic resonance technique to measure the magnetic field $B_{\rm 0}$ \cite{Arnold1947,Bloch1948}. 
In 1956, Cohen, Corngold and Ramsey achieved an uncertainty of $25$~ppm, 
using Ramsey's resonance technique of separated oscillating fields  \cite{Cohen1956}. 
Then, challenged by the possible discovery of a nonzero neutron electric dipole moment (nEDM), Ramsey's technique was further developed. 
Profiting from these developments, Greene and coworkers \cite{Greene1979} 
measured in 1977 the neutron-to-proton magnetic moment ratio with an improvement of two orders of magnitude in accuracy. 
In the latter experiment, the separated field resonance technique was applied simultaneously to a beam of slow neutrons and a flow of protons in water precessing in the same magnetic field. 

Since 1986 the neutron gyromagnetic ratio has been considered by the Committee on Data for Science and Technology (CODATA) as a fundamental constant. 
In the 2010 evaluation of the fundamental constants \cite{CODATA2010}, 
the accepted value 
\begin{equation}
\label{litt_gamma_n}
\frac{\gamma_\n}{2 \pi} = 29.1646943(69) \, \rm{MHz/T} \quad [0.24 \, {\rm ppm}]
\end{equation}
was obtained by combining Greene's measurement \cite{Greene1979} of $\gamma_\n/\gamma_p'$ 
with the determination of the shielded proton gyromagnetic ratio in pure water $\gamma_p'$. 

In this letter we report on a measurement of the neutron magnetic moment with an accuracy of better than 1~ppm, 
using an apparatus that was built to search for the neutron electric dipole moment \cite{Baker2011}. 
The experimental method differs in two essential aspects from the one reported in Ref. \cite{Greene1979}. 
First, we use stored ultracold neutrons (UCNs) rather than a beam of cold neutrons. 
Second, the magnetic field is measured using a co-magnetometer based on the nuclear spin precession of $^{199}$Hg atoms. 
The result presented is, in fact, a measurement of the ratio $\gamma_\n/\gamma_\Hg$ of the neutron to mercury $^{199}$Hg magnetic moments. 

Although not as suitable as the proton, the $^{199}$Hg atom can also be used as a magnetic moment standard. 
In 1960 Cagnac measured $\gamma_\Hg/\gamma_p'$ in Kastler's lab using the newly invented Brossel's double resonance method \cite{Cagnac1961}. 
Since the magnetic field was calibrated using nuclear magnetic resonance in water, the mercury magnetic moment was effectively measured in units of the proton magnetic moment. 
Cagnac's result $\gamma_\Hg/\gamma_p' = 0.1782706(3)$ can be combined with the current accepted value of the shielded proton moment ($\gamma_p'/ 2 \pi = 42.5763866(10)$~MHz/T \cite{CODATA2010}) to extract a mercury gyromagnetic ratio of 
\begin{equation}
\label{litt_gamma_Hg}
\frac{\gamma_\Hg}{2 \pi} =  7.590118(13) \, \rm{MHz/T} \quad [1.71 \, {\rm ppm}]. 
\end{equation}

Our result for $\gamma_\n/\gamma_\Hg$ with an accuracy better than 1 ppm provides an interesting consistency check on results (\ref{litt_gamma_n}) and (\ref{litt_gamma_Hg}). 
It confirms the currently accepted value of the neutron magnetic moment, which was inferred from a single experiment \cite{Greene1979}. 
Alternatively, our result can be used to produce a more accurate value for the mercury magnetic moment. 

\section{The measurement with the nEDM spectrometer}
\label{OILL}

\begin{figure}
\centering
\includegraphics[width=0.97\linewidth]{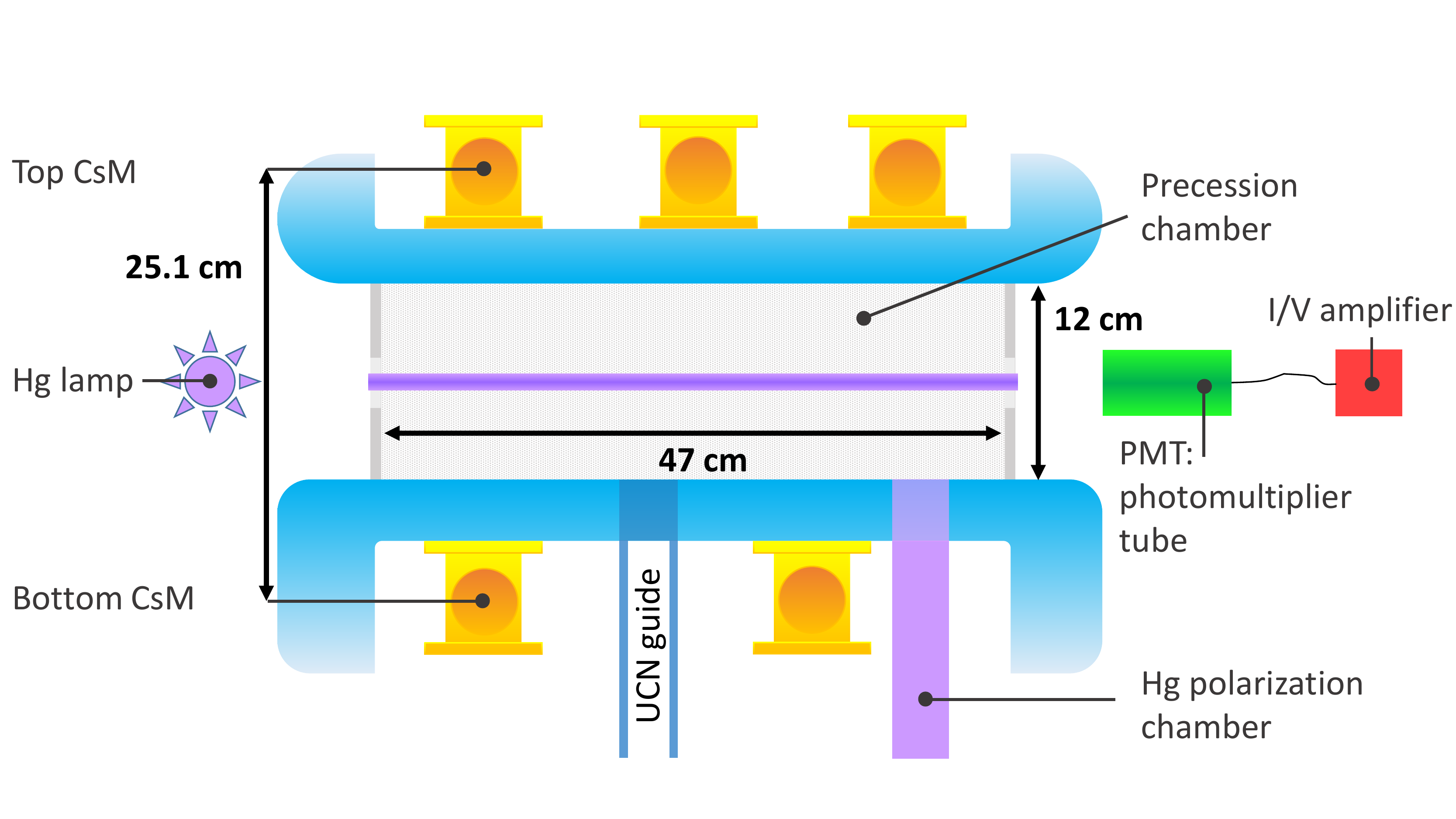}
\caption{
Vertical cut through the inner part of the nEDM apparatus. 
Schematically indicated are the precession chamber, the mercury comagnetometer and the Cs magnetometer array. 
}
\label{scheme}
\end{figure}

The measurement was performed in fall 2012 with the nEDM spectrometer, 
currently installed at the Paul Scherrer Institute (PSI). 
A precursor of this room-temperature apparatus, operated at Institut Laue Langevin (ILL), has produced the currently lowest experimental limit for the neutron EDM \cite{Baker2006}. 
A detailed description of the apparatus, when operated at ILL, can be found in \cite{Baker2014}. 
A schematic view of the core of the apparatus, as installed now at PSI, is depicted in Fig.~\ref{scheme}. 

The spectrometer uses UCNs, neutrons with kinetic energies of $\approx 100$~neV that can be stored in material bottles for a few minutes. 
In a typical measurement cycle, UCNs from the new PSI source \cite{Lauss2014} are guided to the precession chamber, 
filling the volume for 34~s before closing the UCN valve. 
The neutrons are fully polarized on their way from the source to the apparatus by a 5~T superconducting magnet. 
The precession chamber is a 22~liter cylindrical storage volume of height $H=12$~cm made out 
of a deuterated polystyrene ring and two diamond-like-carbon coated flat metallic plates. 
The entire apparatus is evacuated to a residual pressure below $10^{-5}$~mbar during the measurements. 
After the completion of the Ramsey procedure described below, the UCN valve is opened again and the neutrons fall down into a detector where they are counted. 
On the way to the detector they pass through a magnetized iron foil that serves as a spin analyzer, providing a spin-dependent counting of the neutrons. 

The precession chamber is exposed to a static vertical magnetic field of $B_{\rm 0} \approx 1030$~nT, either pointing upwards or downwards, 
corresponding to a neutron Larmor precession frequency of $f_\n = \frac{\gamma_\n}{2 \pi} B_{\rm 0} \approx 30$~Hz. 
This highly homogeneous magnetic field ($\delta B / B \approx 10^{-3}$) is generated by a $\cos \theta$ coil. 
In addition, a set of trim coils permits the optimization of the magnetic field uniformity or the application of magnetic field gradients. 
These coils are wound on the outside of the cylindrical vacuum tank, which is surrounded by a four-layer mu-metal magnetic shield. 

During the storage of polarized UCN, the Ramsey method of separated oscillatory fields is employed  
in order to measure $f_\n$ accurately. 
At the beginning of the storage period, a first oscillating horizontal field pulse of frequency $f_{\rm RF}\approx f_\n$ is applied for $t =2$~s, 
thereby flipping the neutron spins by $\pi/2$. 
Next, the UCN spins are allowed to precess freely in the horizontal plane around the $B_\mathrm{0}$ field, 
for a precession time of $T=180$~s. 
A second $\pi/2$-pulse, at the same frequency $f_{\rm RF}$ and in phase with the first pulse, is then applied.
The Ramsey procedure is resonant in the sense that it flips the neutron spins by exactly $\pi$ only when $f_{\rm RF} = f_\n$. 

In order to monitor the magnetic field $B_{\rm 0}$ within the precession chamber, a cohabiting mercury magnetometer is used \cite{Green1998}. 
A gas of $^{199}$Hg mercury atoms is continuously polarized by optical pumping in a polarization chamber situated below the precession chamber. 
At the beginning of a measurement cycle, the precession chamber is filled with polarized atoms, and a $\pi/2$ mercury pulse is applied immediately before the first neutron pulse. 
Thus, during the precession time, both neutron spins and mercury spins precess in the horizontal plane, sampling the same volume. 
The Larmor frequency $f_\Hg = \frac{\gamma_\Hg}{2 \pi} B_{\rm 0} \approx 8$~Hz of the mercury atoms is measured by optical means: 
 the modulation of the transmission of a circularly polarized resonant UV light beam from a $^{204}$Hg discharge lamp is measured with a photomultiplier tube. 
For each cycle, the mercury comagnetometer provides a measurement of the magnetic field 
averaged over the same period of time as for the neutrons, at an accuracy of $\sigma (f_\Hg) \approx 1 \upmu$Hz. 

\begin{figure}
\centering
\includegraphics[width=0.97\linewidth]{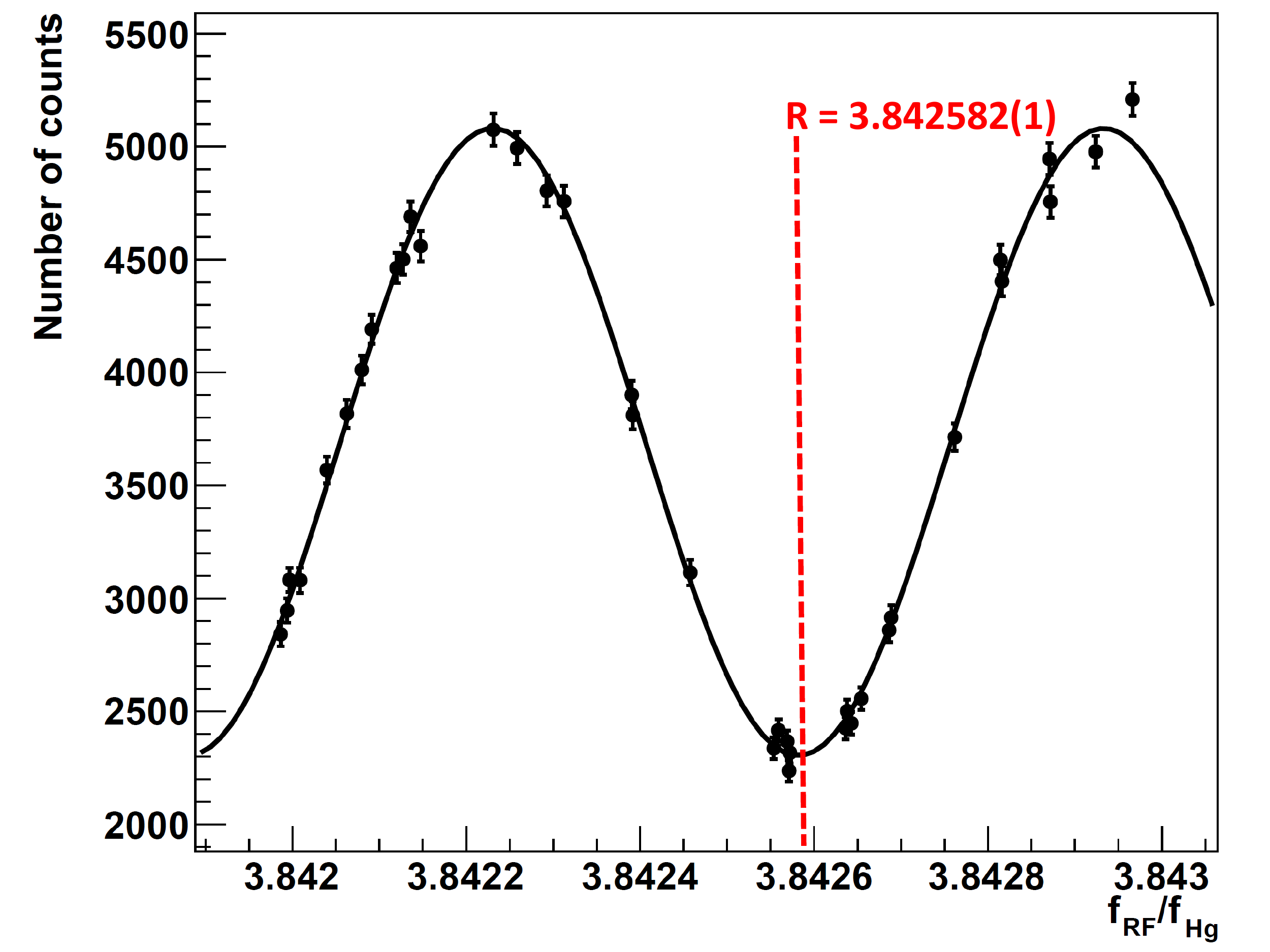}
\caption{Ramsey fit for the run 6043. In this run the goodness-of-fit is quantified by $\chi^2 / d.o.f. = 1.3$. 
}
\label{Run6043_Ramsey}
\end{figure}

A typical run consists of a succession of $\sim$~20 cycles, in which the neutron pulse frequency $f_{\rm RF}$ is randomly varied. 
We analyzed each run to extract the neutron to mercury frequency ratio $R = f_\n/f_\Hg$, which, in absence of systematic effects, 
would be equal to $\gamma_\n / \gamma_\Hg$. 
In a Ramsey experiment, the phase of the spin after the effective precession time $T+4t/\pi$ is $\phi=2 \pi (f_{\rm RF}-f_{\rm n}) (T+4t/\pi)$ (see e.g. \cite{Piegsa2008} for a derivation of the factor $4/\pi$). 
When the magnetic field acting on the neutron is monitored via the mercury precession frequency, then the analysis of a run consists in fitting a Ramsey fringe pattern to the data, according to 
\begin{equation}
N^{\rm up}_i = N_0 \left( 1 - \alpha \cos( K (f_{\rm RF, i}/f_{\Hg, i} - R) ) \right), 
\end{equation}
where $N^{\rm up}_i$ is the number of neutron counts in cycle $i$, $f_{\rm RF, i}$ is the frequency of the neutron $\pi/2$-pulses for cycle $i$ and $f_{\Hg, i}$ the measured mercury precession frequency. 
The parameter $K$ defined by 
\begin{equation}
K = 2 \pi (T + 4 t /\pi) \langle f_\Hg \rangle 
\end{equation}
represents the period of the fringes, 
where $\langle f_\Hg \rangle$ is the mean mercury frequency during the run. 
For each run the three parameters  $\alpha$ (visibility of the Ramsey fringe), $N_0$ and $R$ are extracted from the fit. 
An example of a run is shown in Fig.~\ref{Run6043_Ramsey}. 
The vertical error bars follow from counting statistics and the horizontal error bars are negligible. 
The sensitivity of the technique is thus limited by neutron counting statistics and was on average $\sigma(R) = 1.5 \times 10^{-6}$ per run. 

Different runs correspond to different trim coil current settings and thus different magnetic field gradients. 
The measurements with different magnetic field configurations serve to correct the systematic effects described in the next section. 
For the sake of simplicity we have retained for the final analysis only runs with gradients smaller than $200$~pT/cm, i.e. 7 runs with the magnetic field pointing downwards and 9 runs with the magnetic field pointing upwards. 

\section{Systematic effects}
\label{Systematics}

There are four known effects that could significantly shift the ratio $R = f_\n / f_\Hg$ from its unperturbed value $\gamma_\n/\gamma_\Hg$, either by affecting the neutron frequency or the mercury frequency. 
We write the combination of all these effects as 
\begin{equation}
\label{sysList}
R = \frac{f_\n}{f_\Hg} = \frac{\gamma_\n}{\gamma_\Hg} \left( 1 + \delta_{\rm Grav} + \delta_{\rm T} + \delta_{\rm Light} + \delta_{\rm Earth}   \right), 
\end{equation}
and address them in detail in what follows.

The most important effect is known as the gravitational shift $\delta_{\rm Grav}$. 
The room temperature gas of mercury atoms fills the storage volume with a uniform density in all spatial directions. 
Conversely, the much colder UCN gas is strongly affected by gravity and the UCN density is significantly higher at the bottom 
of the storage chamber than at the top. 
This results in a difference $h$ of the vertical locations of the centers of mass of the two species which, in turn, results in a shift of the ratio $R$ in the presence of a vertical magnetic field gradient: 
\begin{equation}
\delta_{\rm Grav}^{\uparrow/\downarrow} = \pm \frac{h}{B_{\rm 0}} \frac{\partial B}{\partial z} ,
\end{equation}
where the arrows and the $\pm$ sign refer to the direction of the magnetic field. 
A direct measurement of this effect was reported in \cite{Altarev2011}. 

To correct for the gravitational shift, 
the magnetic field gradient was measured using an array of cesium magnetometers (four magnetometers on the top of the chamber and seven on the bottom; see Fig.~\ref{scheme}). 
The working principle of these magnetometers is described in \cite{Knowles2009}. 
The transverse components of the magnetic field are small, therefore the scalar magnetometers are effectively measuring the longitudinal component: $B \approx B_z$. 
For each run, a field value was extracted for each magnetometer by averaging the magnetometer readings over the duration of the run. 
A second-order parameterization of the field was then fitted to the field measured at the positions of the eleven magnetometers. 
The parameterization 
\begin{eqnarray}
\label{param}
& B(x,y,z) =  & b_0 + g_x x + g_y y +g_z z + g_{xx} (x^2-z^2) +  \\ 
\nonumber
& & g_{yy} (y^2-z^2) + g_{xy} xy + g_{xz} xz + g_{yz} yz, 
\end{eqnarray}
is the most general second order polynomial satisfying constraints imposed by Maxwell's equation. 
The nine parameters were extracted by performing an unweighted least-squares minimization. 
The main parameter of interest is $g_z$, since the differences of the UCN and mercury volume averages for the other terms are negligible. 
For an error evaluation on $g_z$, the jackknife procedure was applied, 
by removing one magnetometer at a time and performing the fit always with ten magnetic field values. 
The standard deviation of those eleven values for $g_z$ was then used as the uncertainty on $g_z$. 
The jackknife error obtained by this procedure was $8$~pT/cm. 
This analysis method was tested using toy data \cite{Helaine2014}, 
it was found that the jackknife procedure accounts for the incompleteness of the second order parametrization (\ref{param}) and for the possible offsets of the magnetometers. 

\begin{figure}
\centering
\includegraphics[width=0.9\linewidth,angle=90]{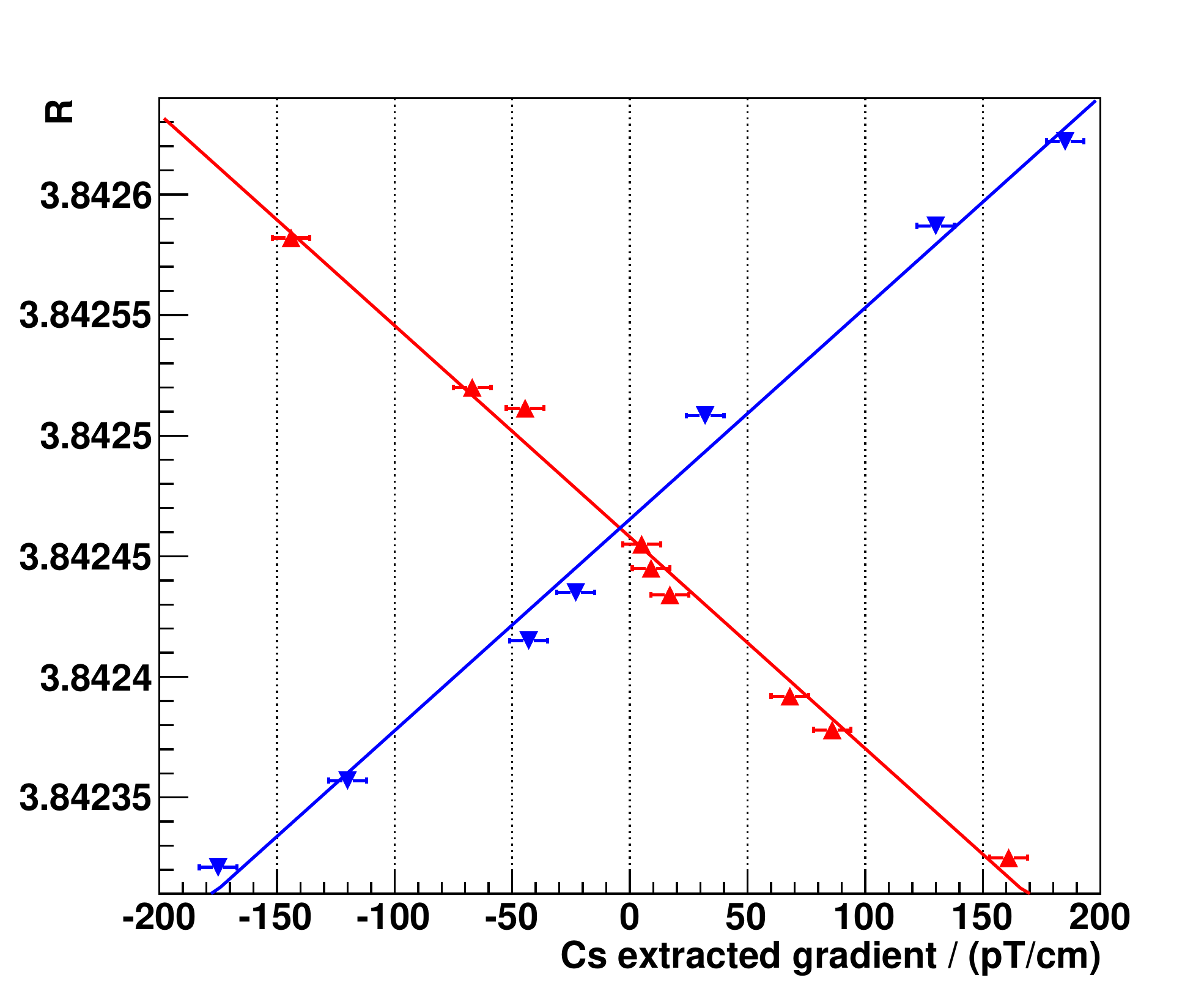}
\caption{
Neutron to mercury frequency $R=f_\n/f_\Hg$ plotted as a function of the field gradient. 
Upwards-pointing (red) and downwards-pointing (blue) triangles represents runs in which $B_{\rm 0}$ was pointing upwards and downwards, respectively. 
The vertical error bars are smaller than the size of the triangles. 
}
\label{Rcurve}
\end{figure}

Figure \ref{Rcurve} shows the $R$ value for each run as a function of the gradient $g_z$. 
We extrapolate the $R$ value to the limit of vanishing gradient by performing a combined fit of the data according to 
\begin{equation}
R^\uparrow   = R_0^\uparrow \left(1 + \frac{h}{B_{\rm 0}} g_{z}^\uparrow \right) \quad {\rm and} \quad 
R^\downarrow = R_0^\downarrow \left(1 - \frac{h}{B_{\rm 0}} g_{z}^\downarrow \right), 
\end{equation}
and obtain the following values for the three fit parameters: 
\begin{eqnarray}
h & = & -0.235(5) \, {\rm cm}, \\
R_0^\uparrow   & = & 3.8424580(23), \\ 
R_0^\downarrow & = & 3.8424653(27). 
\end{eqnarray}
The goodness-of-fit is quantified by $\chi^2 / {\rm d.o.f} = 23/13$, which supports the validity of the gradient errors. 
An analysis of an extended set of data, including runs with higher magnetic field gradients, has been performed independently \cite{Franke2013}, confirming the result presented here. 
The values $R_0^\uparrow$ and $R_0^\downarrow$ are intermediate quantities for which Eq. (\ref{sysList}) holds with the $\delta_{\rm Grav}$ term set to zero. 
The other three systematic effects, which depend solely at first order on the magnetic field direction, still need to be corrected. 

The second systematic effect $\delta_{\rm T}$ arises from residual transverse magnetic field components $B_{\rm T}$. 
The neutron magnetometer is sensitive to the volume average of the magnetic field modulus, viz. $f_\n \propto \langle | \vec{B}| \rangle$ since the neutron Larmor frequency is larger than the wall collision frequency (adiabatic regime). 
On the other hand, the mercury Larmor frequency is much lower than the wall collision frequency. 
As a consequence the mercury magnetometer is sensitive to the vectorial volume average of the field: $f_\Hg \propto | \langle \vec{B} \rangle |$. 
From this essential difference the resulting frequency shift due to a transverse field is given by 
\begin{equation}
\delta_{\rm T} = \frac{\langle {B_{\rm T}}^2 \rangle}{2 B_{\rm 0}^2}, 
\end{equation}
where $\langle {B_{\rm T}}^2 \rangle$ is the volume average over the precession chamber of the squared transverse field. 

To correct for the transverse shift we performed an extensive magnetic-field mapping of the precession chamber region using a three-axis fluxgate magnetometer attached to a robot mapper. 
From the magnetic-field maps we were able to extract the transverse field components in both configurations $B_{\rm 0}$ up and $B_{\rm 0}$ down: 
\begin{equation}
\langle {B_{\rm T}}^2 \rangle^\uparrow = (2.1 \pm 0.5) \, {\rm nT}^2 \; ; \;
\langle {B_{\rm T}}^2 \rangle^\downarrow = (1.7 \pm 0.7) \, {\rm nT}^2. 
\end{equation}
From this we infer corrections given by
\begin{equation}
\delta_{\rm T}^\uparrow = (1.0 \pm 0.2)\times 10^{-6} \; ; \; 
\delta_{\rm T}^\downarrow = (0.8 \pm 0.3)\times 10^{-6}. 
\end{equation}

The third systematic effect $\delta_{\rm Light}$ arises from a possible light shift of the mercury precession frequency induced by the light beam that detects the Hg free-induction decay. 
This phenomenon, discovered in 1961 by Cohen-Tannoudji \cite{CohenTannoudji1961}, 
is a shift of the resonance frequency proportional to the UV light intensity. 
We performed a dedicated test to quantify this effect, by measuring the mercury frequency while varying the light intensity. 
The light shift has a vectorial component, i.e. it could depend on the angle between the light beam axis and the magnetic field direction. 
Therefore the associated frequency shift could in principle depend on the $B_{\rm 0}$ direction. 
We performed the test for both $B_{\rm 0}$ polarities and we report the result: 
\begin{equation}
\delta_{\rm Light}^\uparrow   = (0.34 \pm 0.18)\times 10^{-6} \; ; \; 
\delta_{\rm Light}^\downarrow = (0.21 \pm 0.14)\times 10^{-6}. 
\end{equation}

The last systematic effect $\delta_{\rm Earth}$ is a shift due to the Earth's rotation (see \cite{Lamoreaux2007} for a discussion of this effect in the context of the nEDM). 
The precession frequencies of the neutron and the mercury results 
from the Larmor frequencies in a non-rotating frame combined with the rotation of the Earth. 
One can derive the following expression for the associated frequency shift 
\begin{equation}
\label{earth}
\delta_{\rm Earth}^{\uparrow/\downarrow} = \mp \left( \frac{f_{\rm Earth}}{f_\n} + \frac{f_{\rm Earth}}{f_\Hg} \right) \sin(\lambda) = \mp 1.4 \times 10^{-6}, 
\end{equation}
where $f_{\rm Earth} = 11.6 \, \upmu$Hz is the Earth's rotation frequency and $\sin(\lambda) = 0.738$ the sine of the latitude of the PSI location. 
The conventions are such that $f_\n$ and $f_\Hg$ are positive frequencies. 
In the derivation of Eq. (\ref{earth}) it was important to 
consider that the true neutron and mercury gyromagnetic ratios are negative and positive quantities, respectively.

The relative difference between the $R$ ratios measured with $B_{\rm 0}$ up and $B_{\rm 0}$ down, after correcting for the gravitational, transverse, and light shifts, 
should amount to the Earth's rotation effect, whose anticipated value is $\delta_{\rm Earth}^\uparrow - \delta_{\rm Earth}^\downarrow = -2.7\times 10^{-6}$. 
Indeed we find
\begin{equation}
2 \frac{R_0^\uparrow - R_0^\downarrow}{R_0^\uparrow + R_0^\downarrow} - (\delta_{\rm T}^\uparrow - \delta_{\rm T}^\downarrow) - (\delta_{\rm Light}^\uparrow - \delta_{\rm Light}^\downarrow) = (-2.2 \pm 1.0) \times 10^{-6}, 
\end{equation}
in agreement with the expected value. 

Two additional minor systematic effects were considered and neglected. 
First, the Bloch-Siegert shift of the neutron resonance, associated with the use of a linearly oscillating 
rather than rotating field for applying the $\pi/2$-pulses, is calculated to be $2 \, \upmu$Hz. 
Second, possible biases induced by the DAQ electronics were investigated. 
A dedicated multifunction electronic module \cite{Bourrion2013} synchronized on an atomic clock serves 
 both to generate the neutron pulses and to sample the mercury precession signal. 
We checked that effects that could modify the frequency ratio, such as a loss of phase coherence between the two neutron pulses, are negligible. 

\begin{table}
\center
\caption{
Error budget for the measurement of $\gamma_\n/\gamma_\Hg$}
\begin{small}
\begin{tabular}{l|l|l}
Effect	& $B_{\rm 0} \uparrow$ & $B_{\rm 0} \downarrow$ \\
\hline 
Counting statistics			 & $\pm 0.5 \times 10^{-6}$	  & $\pm 0.5 \times 10^{-6}$ \\
Gravitational shift 		 	 & \multirow{2}{*}{$(-8.9 \pm 2.3) \times 10^{-6}$} & \multirow{2}{*}{$(-1.8 \pm 2.7) \times 10^{-6}$} \\
\hfill ($3.84\times \delta_{\rm Grav}$)  &  	&  \\
\hline
Intermediate $R_0$			& $3.8424580(23)$ & $3.8424653(27)$ \\
\hline
Transverse shift 	 		 &  \multirow{2}{*}{$(3.7\pm0.8)\times10^{-6}$}	& \multirow{2}{*}{$(3.0\pm1.2)\times10^{-6}$} \\
\hfill ($3.84\times \delta_{\rm T}$) 	 &  	& \\
Light shift 		 		 & \multirow{2}{*}{$(1.3 \pm 0.7) \times 10^{-6}$} & \multirow{2}{*}{$(0.8 \pm 0.6) \times 10^{-6}$} \\
\hfill ($3.84\times \delta_{\rm Light}$) &  	&  \\
Earth rotation 			 	 & \multirow{2}{*}{$-5.3\times10^{-6}$}	& \multirow{2}{*}{$+5.3\times10^{-6}$} \\
\hfill ($3.84\times \delta_{\rm Earth}$) & 	& \\
\hline
Corrected value				& $3.8424583(26)$	& $3.8424562(30)$ \\
\hline
Combined final $\gamma_\n/\gamma_\Hg$	& \multicolumn{2}{c}{$3.8424574(30)$}
\end{tabular}
\end{small}
\label{budget}
\end{table}

\begin{table}
\center
\caption{
Key data for each run selected for the analysis: 
the fitted visibility $\alpha$, the ratio $R$, the average field $B_{\rm Hg}$ extracted from the mercury comagnetometer, the gradient $g_z$ extracted from the Cs magnetometer array and the squared transverse field $\langle {B_{\rm T}}^2 \rangle$ extracted from the field maps are reported. 
}
\begin{small}
\begin{tabular}{llllll}
Run \#  & $\alpha$ & $R$     & $B_{\rm Hg}$ & $g_z$    & $\langle {B_{\rm T}}^2 \rangle$ \\
        &          &         & [nT]         & [pT/cm]  & [nT$^2$] \\
\hline
\multicolumn{6}{c}{$B_{\rm 0} \downarrow$} \\
\hline
6015   & 0.41 & 3.842321 & 1031.86       & -175      & 2.5 \\
6016   & 0.48 & 3.842587 & 1031.33       & 130       & 1.8 \\
6023   & 0.66 & 3.842435 & 1031.60       & -23       & 1.1 \\
6027-8 & 0.57 & 3.842508 & 1030.18       & 32        & 1.0 \\
6030   & 0.55 & 3.842415 & 1030.32       & -43       & 1.0 \\
6031   & 0.38 & 3.842622 & 1029.92       & 185       & 2.6 \\
6033   & 0.48 & 3.842357 & 1030.45       & -120      & 1.7 \\
\hline
\multicolumn{6}{c}{$B_{\rm 0} \uparrow$} \\
\hline
6040-1 & 0.54 & 3.842445 & 1027.97       & 9        & 1.7 \\
6042   & 0.36 & 3.842325 & 1028.23       & 161      & 2.6 \\
6043   & 0.38 & 3.842582 & 1027.70       & -144     & 3.0 \\
6047   & 0.46 & 3.842520 & 1027.82       & -67      & 2.1 \\
6049   & 0.43 & 3.842378 & 1028.14       & 86       & 1.9 \\
6058   & 0.53 & 3.842434 & 1027.82       & 17       & 1.7 \\
6059   & 0.42 & 3.842511 & 1026.82       & -44      & 2.0 \\
6060   & 0.55 & 3.842455 & 1028.25       & 5        & 1.7 \\
6064   & 0.44 & 3.842392 & 1029.47       & 68       & 1.9
\end{tabular}
\end{small}
\label{keydata}
\end{table}

The error budget and the correction procedure are summarized in Table \ref{budget} and the key data for each run is presented in Table \ref{keydata}. 
As a final step we combine the results obtained for $B_{\rm 0}$ up and $B_{\rm 0}$ down. 
The agreement between these two results provides a nontrivial consistency check. 
To avoid double use of data, we conservatively quote for the final uncertainty the largest of the two individual errors: 
\begin{equation}
\gamma_\n / \gamma_\Hg = 3.8424574(30) \quad [0.78 \, {\rm ppm}]. 
\label{FinalR}
\end{equation}

\section{Discussion}
\label{discussion}

\begin{figure}
\centering
\includegraphics[width=0.97\linewidth]{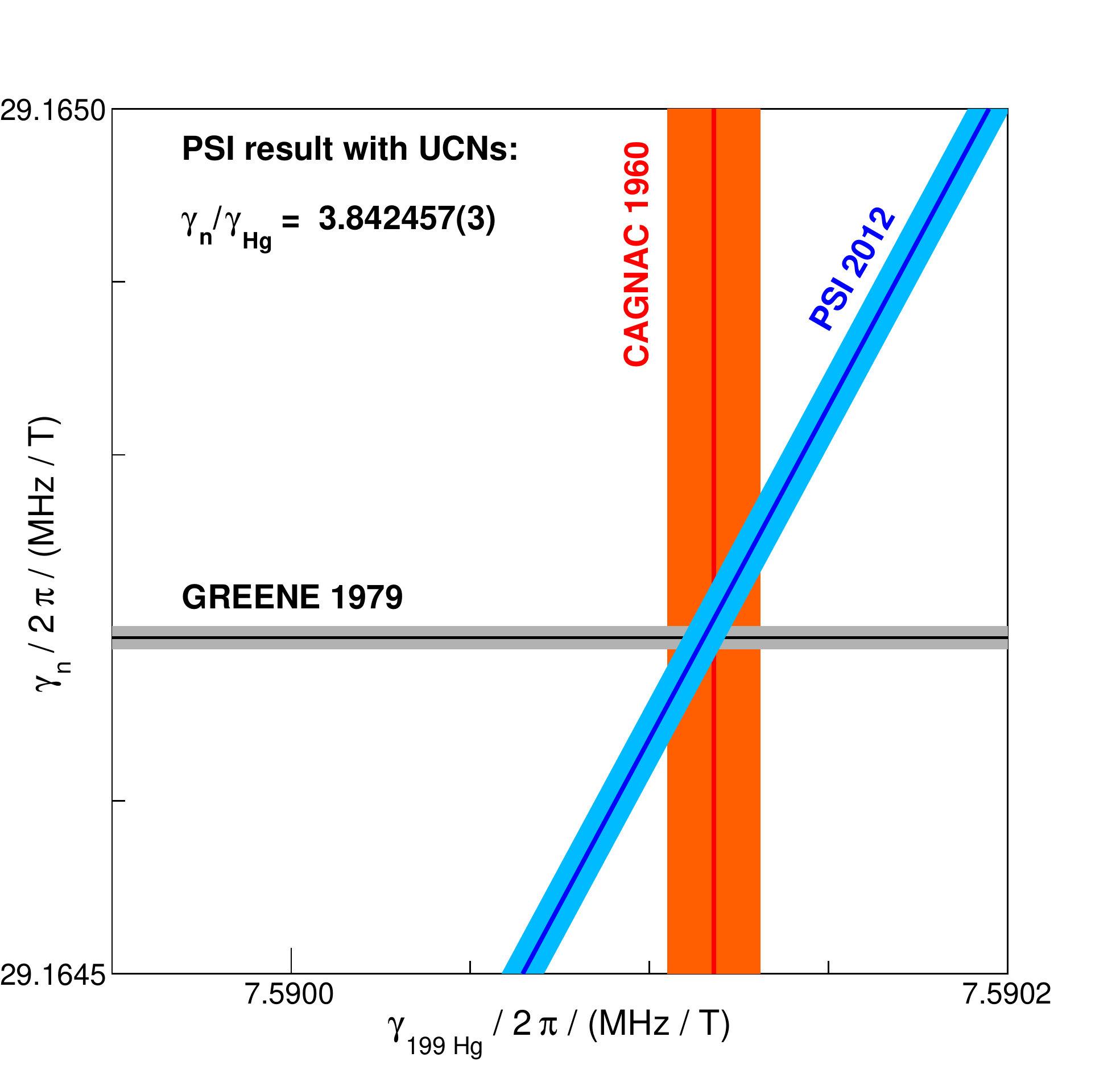}
\caption{
1-sigma allowed regions in the $\gamma_\n, \gamma_\Hg$ plane. 
Our final value for the neutron to mercury magnetic moment ratio (\ref{FinalR}) here labeled as "PSI 2012" forms the diagonal band. 
The horizontal band is the neutron magnetic moment (\ref{litt_gamma_n}) value from Greene et al. and the vertical band is from the measurement of the mercury magnetic moment (\ref{litt_gamma_Hg}) by Cagnac. 
}
\label{final}
\end{figure}

Our result (\ref{FinalR}) can be considered as a consistency check of the accepted values for the neutron and $^{199}$Hg magnetic moments as shown in Fig.~\ref{final}. 
Equivalently, this result provides a new accurate and independent measurement of the neutron gyromagnetic ratio. 
Using (\ref{litt_gamma_Hg}), we obtain:
\begin{equation}
\label{PSI_gamma_n}
\frac{\gamma_\n}{2 \pi} = 29.164705(55) \, \rm{MHz/T} \quad [1.89 \, {\rm ppm}], 
\end{equation}
thus confirming the accepted value which, until now, was based on a single measurement \cite{CODATA2010}. 

Alternatively, we can use the results (\ref{litt_gamma_n}) and (\ref{FinalR}) to propose a new, more accurate value of the $^{199}$Hg atomic gyromagnetic ratio:
\begin{equation}
\label{PSI_gamma_Hg}
\frac{\gamma_\Hg}{2 \pi} = 7.5901152(62) \, \rm{MHz/T} \quad [0.82 \, {\rm ppm}].
\end{equation}

\section*{Acknowledgments}
We are grateful to the PSI staff (the accelerator operating team and the BSQ group) for providing excellent running conditions and acknowledge the outstanding support of M. Meier and F. Burri. 
Support by the Swiss National Science Foundation Projects 200020-144473, 200021-126562, 200020-149211 (PSI) and 200020-140421 (Fribourg) is gratefully acknowledged. 
The LPC Caen and the LPSC acknowledge the support of the French Agence Nationale de la Recherche (ANR) under reference ANR-09--BLAN-0046. 
Polish partners want to acknowledge The National Science Centre, Poland, for the grant No. UMO-2012/04/M/ST2/00556. 
This work was partly supported by the Fund for Scientific Research Flanders (FWO), and Project GOA/2010/10 of the KU Leuven. 
The original apparatus was funded by grants from the UK’s PPARC.

\end{document}